\documentstyle[epsfig,aps,prl,amssymb]{revtex}
\renewcommand{\baselinestretch}{2.}

\begin{document}
\title{New constraints on the mass composition of cosmic rays above  $10^{17}$ eV from Volcano Ranch measurements}
\author{M. T. Dova, M. E. Mance\~nido, A. G. Mariazzi}
\address{Instituto de  F\'{\i}sica, CONICET, Dto. de F\'{\i}sica, Universidad Nacional de La Plata, C.C.67, 1900\\
La Plata, Argentina}
\author{T. P. McCauley}
\address{Department of Physics, Northeastern University, Boston, MA 02115, USA}
\author{A. A. Watson}
\address{Department of Physics and Astronomy, University of Leeds, Leeds, LS2 9JT, UK }

\maketitle

\begin{abstract}
\\
Linsley used the Volcano Ranch array to collect data 
on the lateral distribution of showers produced by cosmic rays at energies 
above $10^{17}$ {\rm eV}. Very precise measurements of the steepness of 
the lateral distribution function were made on 366 events. 
The current availability of sophisticated hadronic interaction models has prompted an interpretation of the measurements. 
In this analysis we use the {\sc aires} Monte Carlo code to 
generate showers, together with {\sc geant4} to simulate the detector response 
to ground particles. 
The results show that, with the assumption of a bi-modal proton and iron mix, 
iron is the dominant component of cosmic rays between $5\times10^{17}$ and $10^{19}$  eV, assuming that hadronic interactions are well-described by {\sc qgsjet} at this energy range. 
\end{abstract}


\section{Introduction}
The measurement of the mass composition of cosmic rays above  $10^{17}$ {\rm eV } 
is a challenging problem.
This information is as important as the energy spectrum and the anisotropy in determining 
cosmic ray origin. One must know the likely mass range of a particular data set 
before one can  interpret anisotropy information confidently, given the influence of galactic 
and intergalactic magnetic fields.
Our knowledge of the mass composition of cosmic rays above $10^{17}$ {\rm eV } remains very limited. 
Recent re-interpretation of measurements of the lateral distribution of 
water-\v{C}erenkov signals made at Haverah Park ~\cite{HaverahPark} suggests a composition of
34\% protons and 66\% iron in the range $2\times10^{17}$ - $10^{18}$ eV.
This contrasts with earlier claims, from observations made using Fly's Eye, 
that the composition changes from a heavy mix around $3\times10^{17}$ {\rm eV} to a proton 
dominated flux around $10^{19}$ {\rm eV}~\cite{FlyEye1}. 
At Yakutsk, both the inferred values of the depth of shower maximum ($X_{max}$) and the muon density favor a composition change from a mixture of
 heavy and light components to light composition over the same energy region~\cite{Yakust}.
From HiRes/MIA data~\cite{HiresMia}, there are claims that there is a rapid change 
from a heavy to a light composition between 0.1 and 1.0 EeV.
A recent analysis by the HiRes collaboration of data collected in the energy range between $10^{18}$ - $10^{19.4}$ eV~\cite{Hires2} is consistent with a nearly constant 
purely protonic composition. The fraction of protons, however, 
decreases when they interpret their data using {\sc sibyll2.1}. 
On the other hand, the AGASA group have argued for a ``mixed'' unchanging 
composition from 1 - 10 EeV~\cite{Agasa} (using {\sc mocca} for the simulations).
A recent analysis of the muon component in air showers with
{\sc aires/qgsjet98} around $10^{19}$ eV by the AGASA collaboration 
indicates a relatively light average composition~\cite{Agasa2}.
 
The source of the discrepancy between different experiments is not understood 
and it is important to resolve the issue, because of its implications for 
cosmic ray models of origin, acceleration and propagation. 
Volcano Ranch data may provide a path for further understanding.

Following the successful re-examination of the Haverah Park 
data~\cite{HaverahPark} with modern shower models, we report here 
a similar analysis using the Volcano Ranch data, collected 
by Linsley~\cite{vrall} to determine the shape of the lateral 
distribution of air showers.
This is the first attempt to examine the Volcano Ranch
data with the results of Monte Carlo calculations, using 
Monte Carlo tools that were unavailable when the data were recorded in 1970.  
It is timely as the situation on mass composition above $10^{17}$ eV 
remains confused and the steepness of the lateral 
distribution is sensitive to the depth of maximum of the shower, 
and therefore to the primary composition and to the character of 
the initial hadronic interactions.

To simulate the development of the air showers, we have used 
the {\sc aires} \cite{aires} code (version 2.4.0),
with the hadronic interaction generator {\sc qgsjet98}~\cite{qgsjet}. 
The results of the simulated showers were convolved with a simulation of the 
detector response made using {\sc geant4}~\cite{geant4}.
A comparison of two hadronic generators ({\sc qgsjet98} and {\sc sibyll2.1}
) was presented in~\cite{isvhecri}.
Both give satisfactory descriptions of the data, but we have preferred to use
{\sc qgsjet98} because this model has been shown to be consistent with
experimental data at energies up to 10 PeV and beyond\cite{kascatere,kascaalan}

\section{The Volcano Ranch Array}

\begin{center} 
\begin{figure}[p]
\epsfig{file=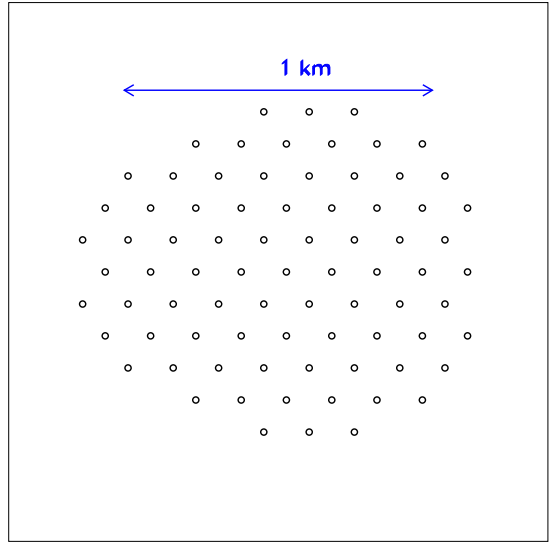,width=0.4\textwidth}
\caption{Volcano Ranch array in the final configuration.}
\label{fig:VRarray}
\end{figure}
\end{center}

The pioneering Volcano Ranch instrument consisted of an array
of scintillation counters. The array was operated in three configurations
from 1959-1976 at the MIT Volcano Ranch station located near Albuquerque,
New Mexico (atmospheric depth 834 g ${\rm cm^{-2}}$). 
One of its many distinctions was the detection of the first
cosmic ray with an energy estimated at $10^{20}${\rm eV}~\cite{vrevent}.
The final configuration, of relevance here, comprised 80 detectors of surface area
0.815 ${\rm m^2}$, scintillator thickness of 9.032 g ${\rm cm^{-2}}$
laid out on a hexagonal grid with a separation of 147 {\rm m} (Fig\ref{fig:VRarray}).
This configuration allowed precise measurement of the lateral
distribution of the detector signals. The steepness of the lateral
distribution, and its fluctuations, can be used to explore the primary
mass composition as in [1]. Fortunately, in his various writings, Linsley has left
unusually detailed descriptions of his equipment, together with examples of events 
and a description of his data reduction methods.

\section{Lateral distribution function}

A generalized version of the Nishimura-Kamata-Greisen (NKG) formula was used 
to describe the lateral distribution of particles at ground in minimum ionizing particles per square metre (${\rm mips\,m^{-2}}$) 
for Volcano Ranch data~\cite{denver}. 
This lateral distribution function is given as 
\begin{equation}
S_{{\rm VR}}(r) = {\frac{N_{fit}}{r_{m}^{2}}} C(\alpha,\eta) \left( {\frac{r}{r_{m}%
}} \right)^{-\alpha} \left( 1 + {\frac{r}{r_{m}}} \right)^{-(\eta-\alpha)}
\end{equation}
normalized to shower size $N_{fit}$ with 
\begin{equation}
C = {\frac{\Gamma(\eta-\alpha)}{2\pi\Gamma(2-\alpha)\Gamma(\eta-2)}} .
\end{equation}

Here $r_m$ is the Moli\`ere radius, which is $\simeq $ 100 {\rm m}
for the Volcano Ranch elevation. $\eta $ and $\alpha $ are parameters that
describe the logarithmic slope of this function.

From a subset of 366 showers detected with the array, the 
form of $\eta $ as a function of zenith angle $%
\theta $ and shower size $N_{fit}$ was found to be~\cite{plovdiv1}: 
\begin{equation}
\langle \eta(\theta, N_{fit}) \rangle = a + b(\sec \theta - 1) +
c\:log_{10}({\frac{N_{fit}}{10^{8}}})
\end{equation}
with $a = 3.88 \pm 0.054$, $b = - 0.64 \pm 0.07$, and $%
c = 0.07 \pm 0.03$ where a fixed value of $\alpha =1$ was adopted.

\section{Simulation of the detector response of the Volcano Ranch array}

The {\sc aires} code provides a realistic air shower simulation system, which includes 
electromagnetic algorithms \cite{hillas2} and links to different hadronic interactions models. 
As mentioned above, we have used the {\sc qgsjet98} model for nuclear fragmentation and
inelastic collisions.
For the highest energy showers, the number of secondaries becomes so large that it is prohibitive 
in computing time and disk space to follow and store all of them. 
Hillas \cite{hillasthi} introduced a non-uniform statistical sampling mechanism
which allows reconstruction of the whole extensive air shower from a small but representative
fraction of secondaries that are fully tracked. Statistical weights are
assigned to the sampled particles to account for the energy of the discarded
particles. This technique is known as ``statistical thinning''. 
The {\sc aires} code includes an extended thinning algorithm, 
which has been explained in detail \cite{aires}. 
The present work has been carried out using, in most cases, an
effective thinning level $\epsilon _{th}=E_{th}/E_{prim}=10^{-7}$ which is
sufficient to avoid the generation of spurious fluctuations
and to provide a statistically reliable sample of particles far from the
shower core. All shower particles with energies above the following
thresholds were tracked: 90 keV for photons, 90 keV for electrons and
positrons, 10 MeV for muons, 60 MeV for mesons and 120 MeV for nucleons and
nuclei. 

\begin{center}
\begin{figure}[p]
\centerline{ 
\epsfig{file=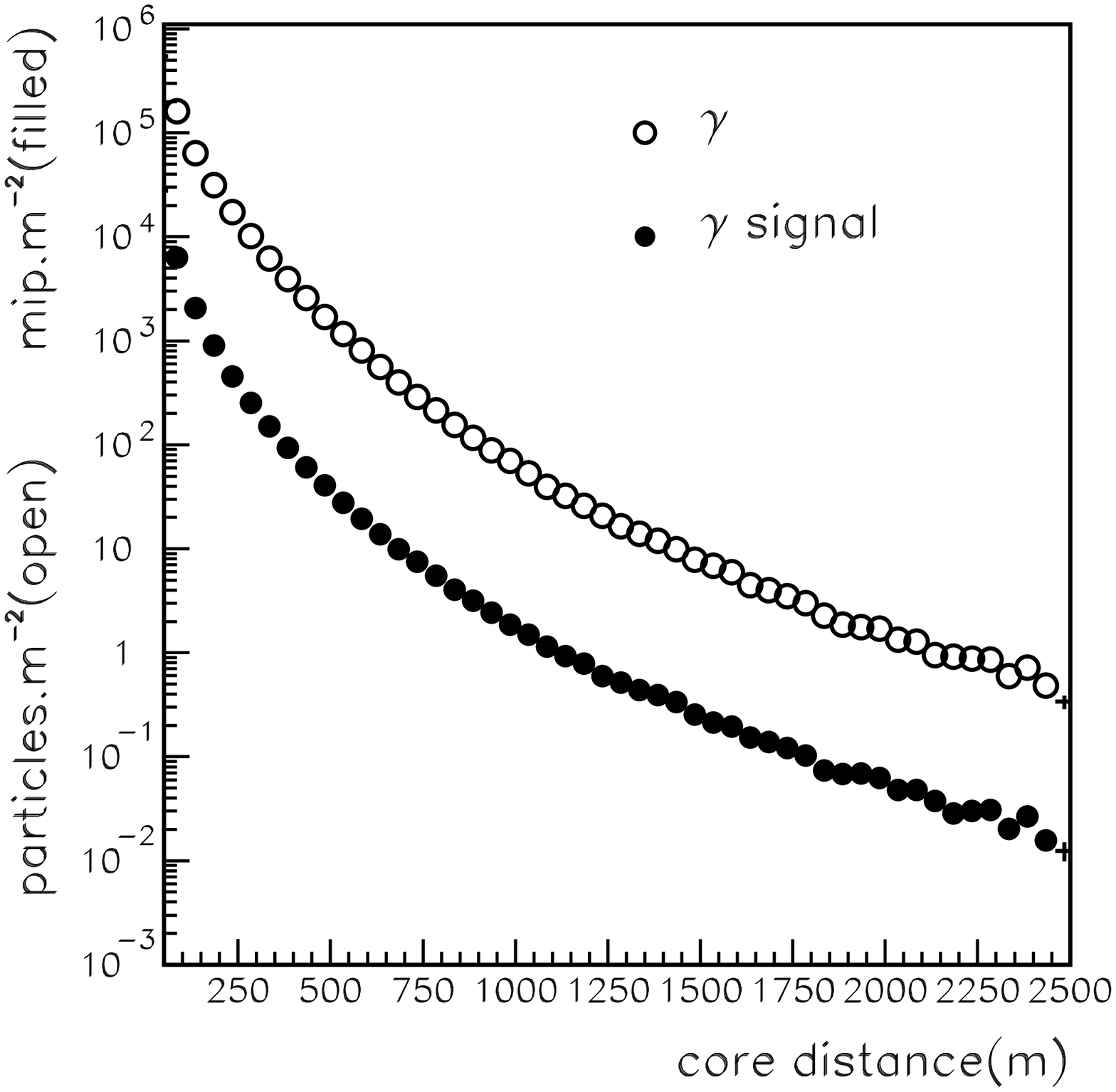,width=6.cm}
\hspace{-12cm}
\epsfig{file=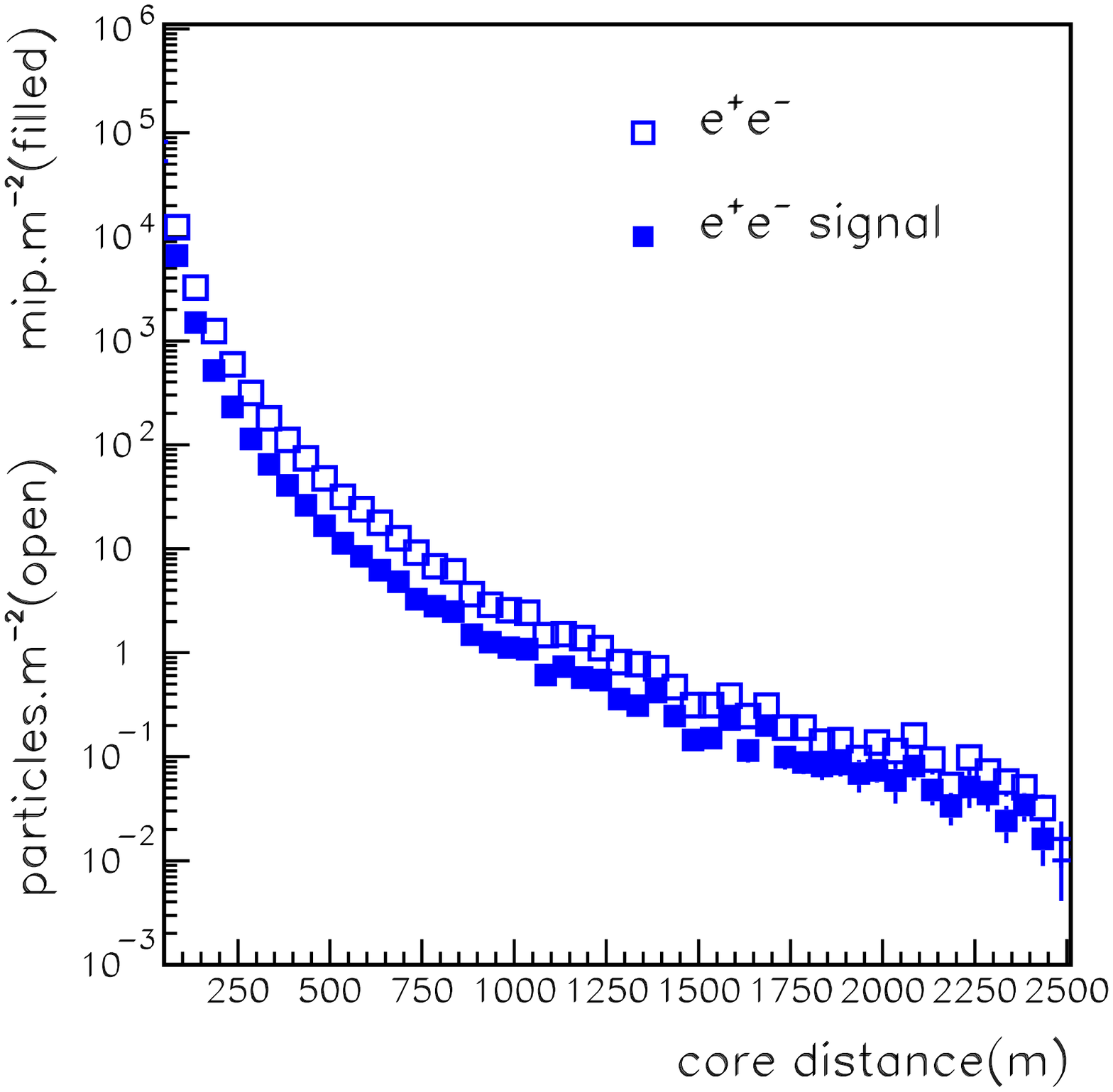,width=6.cm}
\hspace{-12cm}
\epsfig{file=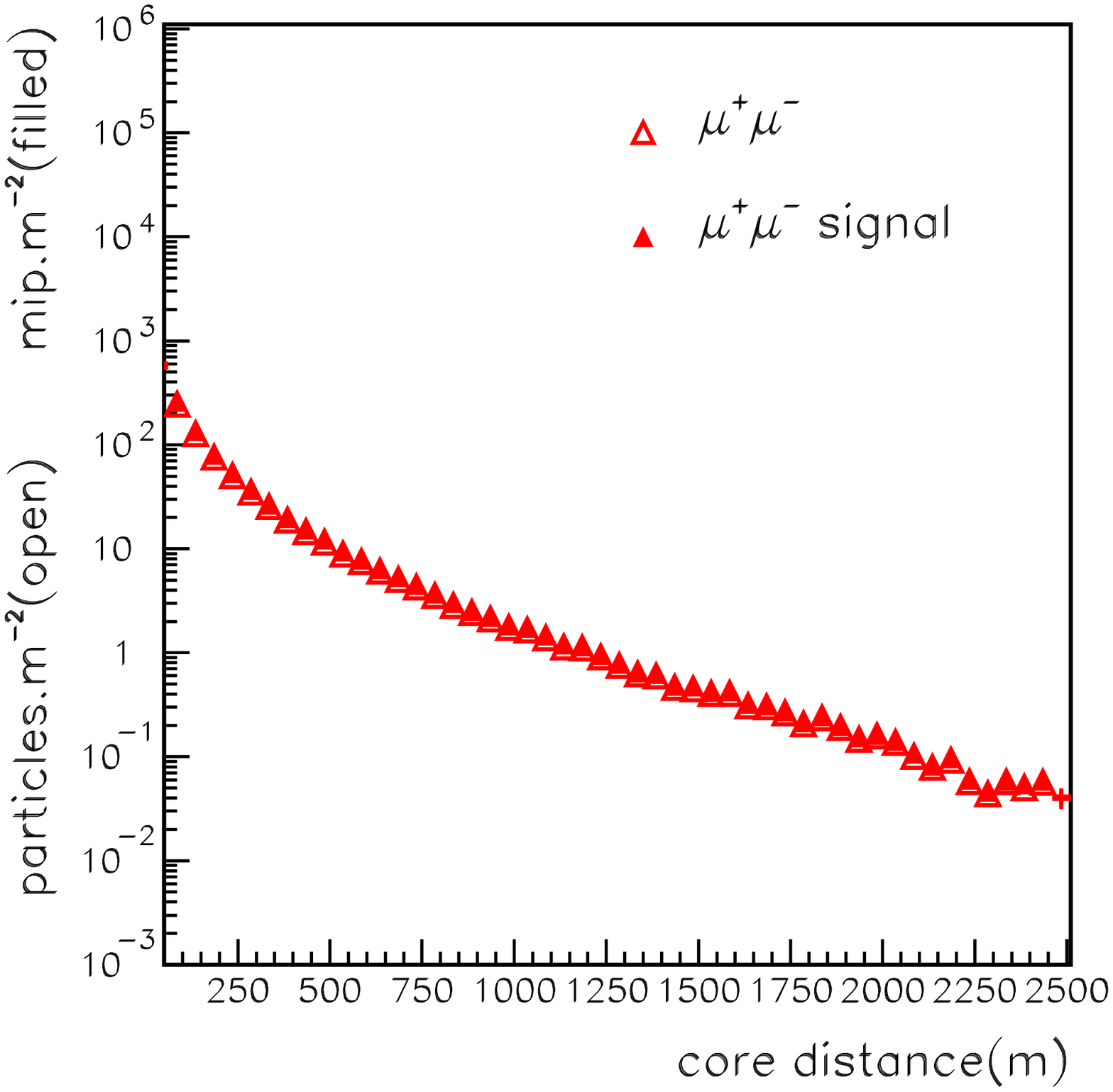,width=6.cm}
}
\caption{Simulated lateral distributions of the three main shower components at ground level and its convolution with the detector response.}
\label{fig:signal}
\end{figure}
\end{center}

We have generated a total of 1735 proton and iron showers with zenith angles 
in the range $\sec {\theta }$ = 1.0 - 1.5 and 
primary energies 
between $10^{17}$ eV and $10^{19}$ eV, to match the Volcano Ranch data. 
To simulate the response of the detectors of the array to the ground
particles, we utilized the general-purpose simulation toolkit {\sc geant4}.
Our procedure follows the prescription in~\cite{kutter}, where the detector
response to electrons, gamma, and muons is simulated in the energy range 0.1
to $10^5$ {\rm MeV} and for five bins per decade of energy. The results of air shower
simulations are convolved with the detector response to obtain 
the scintillator yield expressed in ${\rm mips\ m^{-2}}$. The 
computed lateral distributions of particles and the corresponding 
signal from the scintillators for photons, electrons and muons
are displayed in Figure~\ref{fig:signal}. 
A comparison between the lateral distribution measurements\cite{denver} and 
proton/iron showers simulated with {\sc aires/qgsjet98} including the 
scintillator response was presented previously \cite{isvhecri}. Each simulated shower was thrown a maximum of 100 times on to the  
simulated Volcano Ranch array with random core positions 
in the range 0 - 500 m from the array center.

\begin{center}
\begin{figure}
\centerline{ 
\epsfig{file=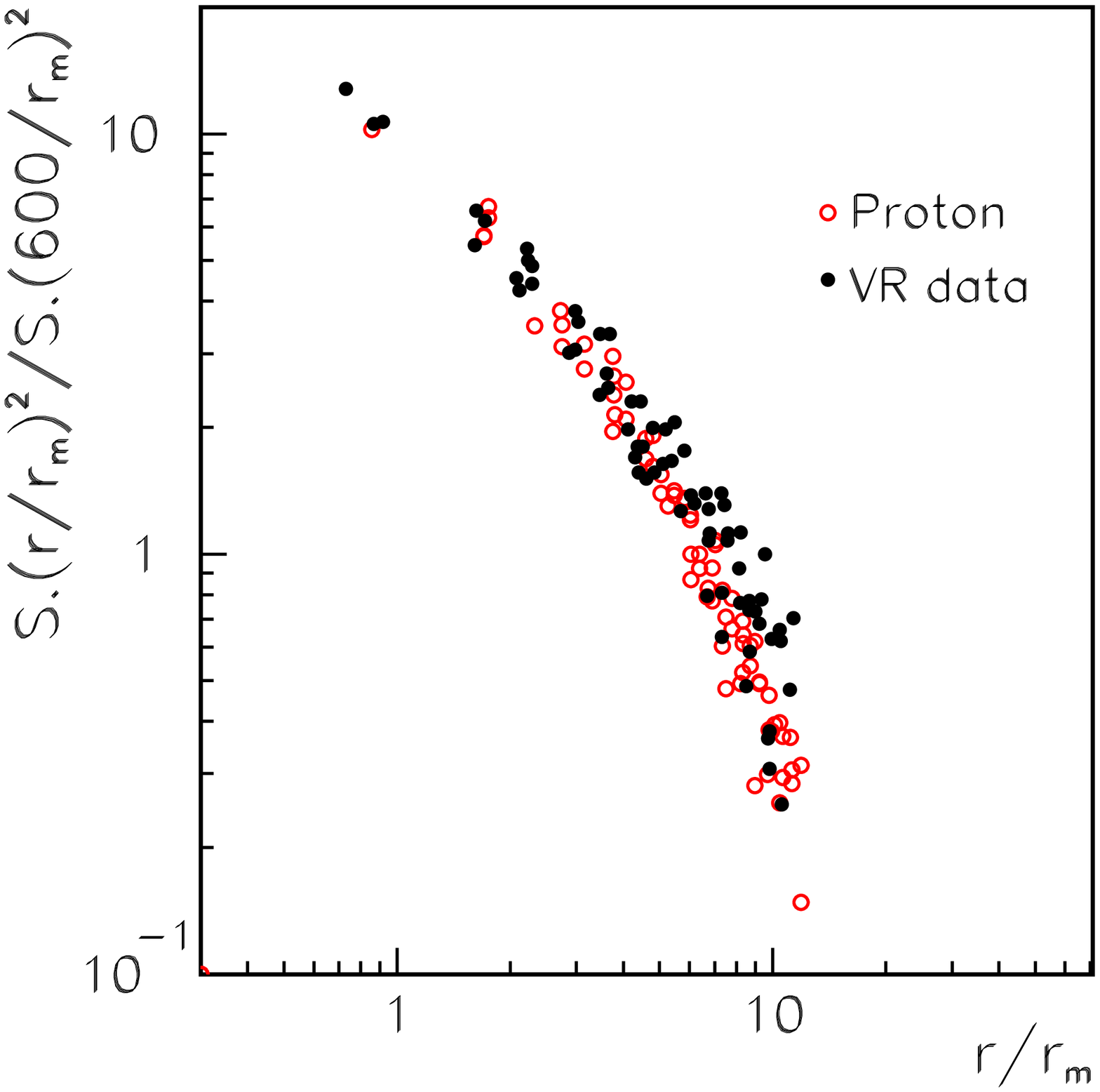,width=8cm}
\hspace{1cm}
\epsfig{file=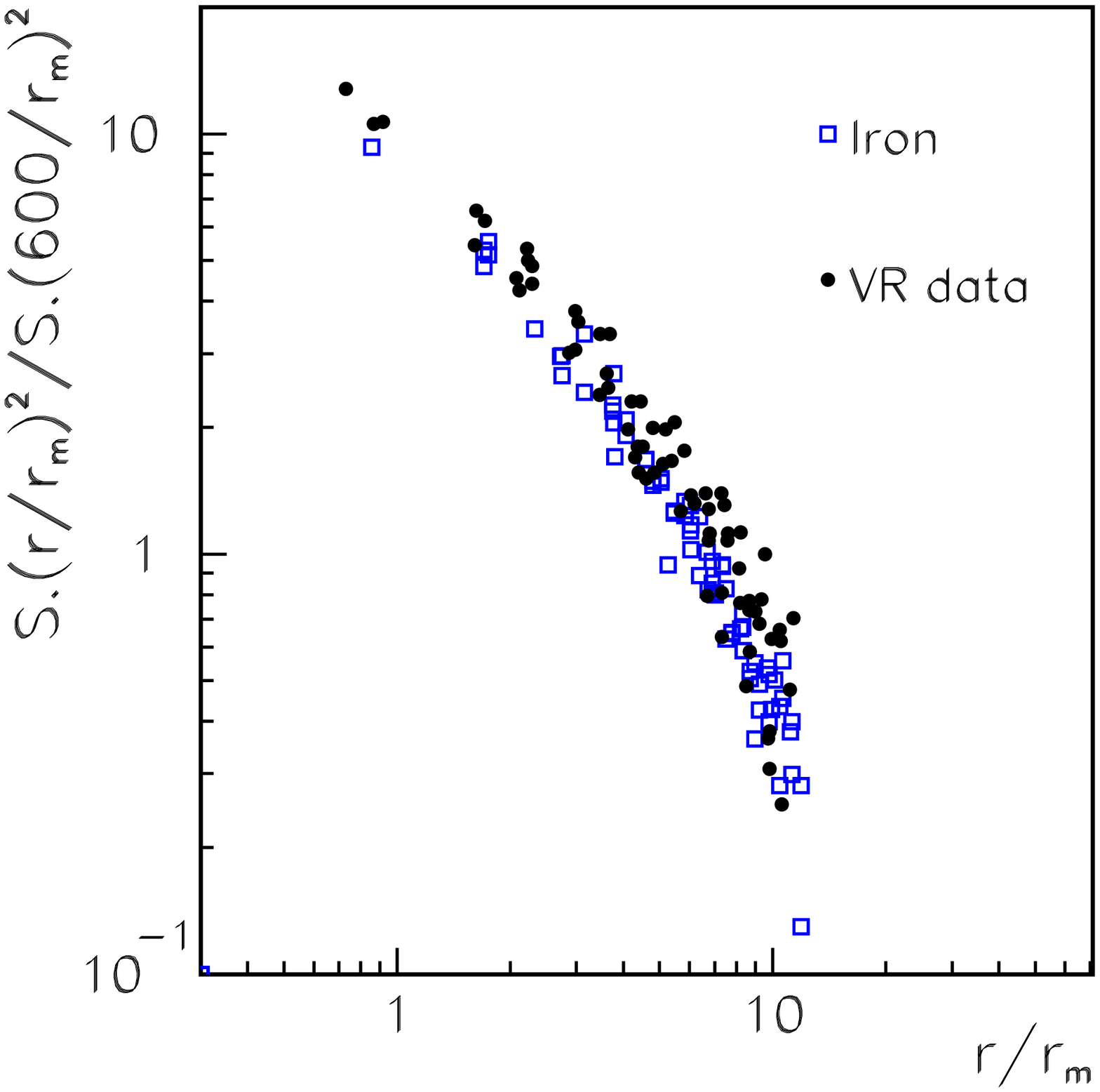,width=8cm}  }
\caption{ Comparison between lateral distribution measurements in a single event [24] and the simulated scintillator response in the configuration of VR array for $10^{19.1}$ eV proton and iron showers.}
\label{fig:array1}
\end{figure}

\end{center}

With the thinning method used in Monte Carlo shower generators, 
when particles reach the thinning energy just one of them is followed 
and multiplied by the corresponding weight at the end. 
Thus, to simulate the response of the detectors correctly, it is necessary to perform 
a smoothing of the densities of the ground particles around the position of each detector.
All particles in a ``sampling zone''  around a given detector are selected and the
statistical weight, as obtained from {\sc aires}, is multiplied by the
``sampling ratio'' $A_{detector}/A_{sampling}$ where $A_{sampling}$ is the area of the ``sampling zone'' and  
$A_{detector}$ is the corresponding detector area. This is equivalent to sampling particles on a larger
area to get a realistic density around the detector position.
As the densities depend mainly on the distance to the shower axis, the sampling area
over which simulated particles are gathered is such that this ratio varies from
about 0.1 at 100 m to about 0.001 at 1 km.
As a first check on the validity of our approach, the data of a single 
large event were compared with calculations~\cite{isvhecri}.
Further checks between data and Monte Carlo were performed as 
the one in Figure~\ref{fig:array1}.
In this plot we present a comparison between lateral distribution 
measurements 
\cite{denver} and a $10^{19.1}$ eV proton (left) and iron (right) shower simulated 
with {\sc aires/qgsjet98}, including the scintillator response of the detectors in the Volcano Ranch
array configuration. 
It has been shown ~\cite{Hillas}
that the fluctuation of the density of shower particles far from the core 
is quite small and that the density at 600 m, S(600m), depends only 
on primary energy.
We normalize the showers to the value of S(600m) in order to decouple 
the normalization factor from the parameters related to the shape of the 
lateral distribution which change with primary mass but only slightly with shower size. The agreement between data and Monte Carlo is good and gives
confidence in the procedures used. 

\section{Derivation of the primary mass composition}
The nature of the primaries that initiate air showers is 
difficult to establish 
from the average properties of the data. For example, an average property 
can be explained with a mass composition of a single species (A) 
or by an appropriate mixture of species.
However, with the Volcano Ranch array, accurate 
measurements of $\eta $ were made on a shower-by-shower basis for fixed bins of zenith angle 
separated by 80 g cm$^{-2}$ ~\cite{plovdiv2}. Thus the 
fluctuations of $\eta$ can be used 
to break this degeneracy.
Linsley determined the precision of each measurement of $\eta $ and reported the average value of 
this quantity for each zenith angle bin. 

The average error in $\eta$ from the fit made to the simulated lateral distributions
($\sigma_{sim}=0.029$), is smaller than the one reported by Linsley($\sigma=0.072$).
To include within the simulation the effect of data reconstruction,
we smeared each value of $\eta$ calculated by Monte Carlo 
using a Gaussian with a width chosen so that Linsley's 
overall uncertainties in $\eta $ are reproduced i.e.
$\sigma_{smear}=0.0662$ is found by quadrative subtraction of the average values of
$\sigma_{sim}$ and $\sigma$.

This is a minor correction since the measurement accuracy is so much smaller 
than the intrinsic shower-to-shower spread (r.m.s.= 0.19).
Thus, for each value of $\eta$ found from the Monte Carlo calculation, we have 
a corresponding and realistic estimate of its ``experimental'' uncertainty. 
We are thus able to make comparisons of Volcano Ranch data with our calculations.

\begin{center}
\begin{figure}[p]
\centerline{ 
\epsfig{file=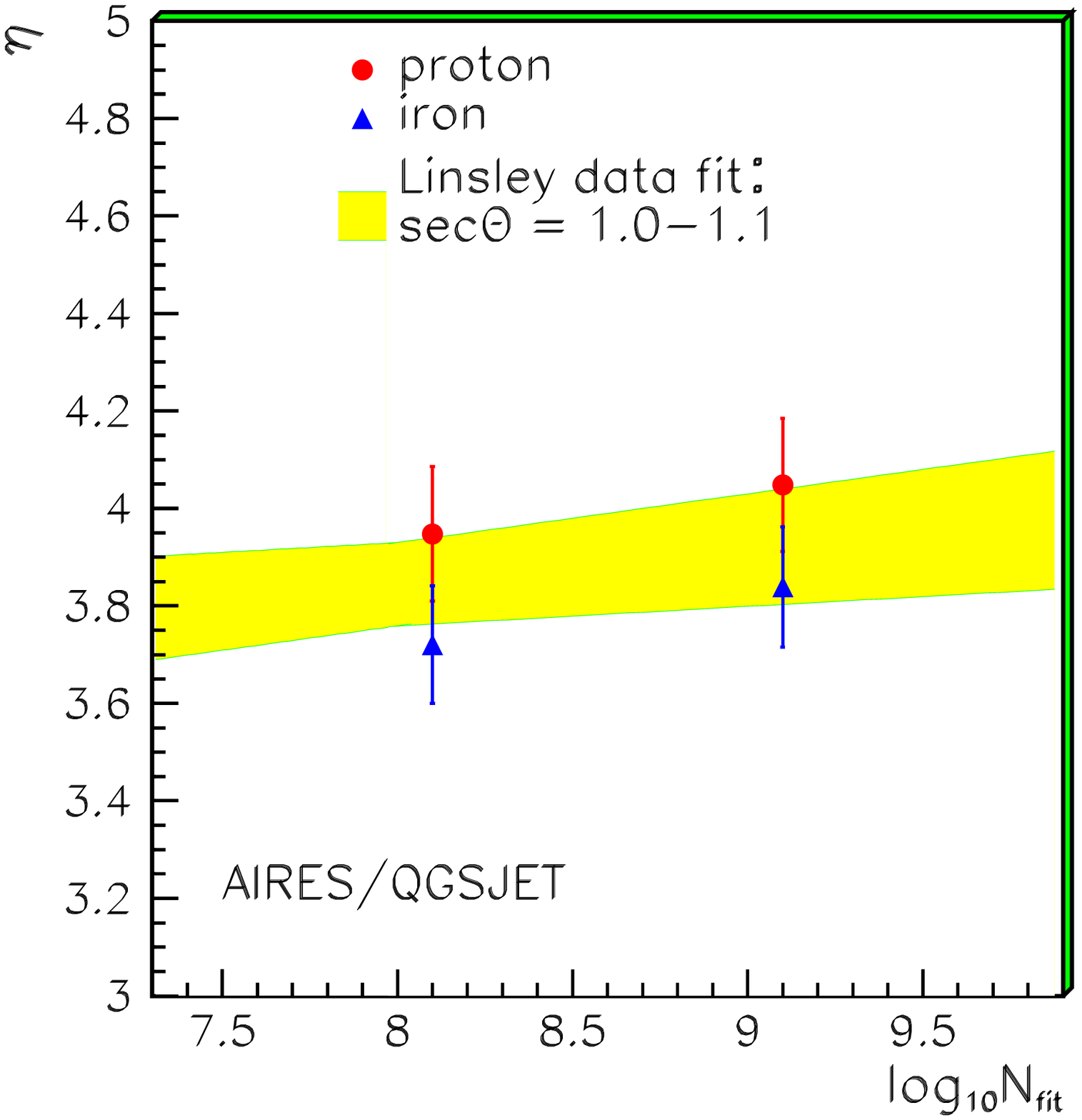,width=8cm}
\hspace{1cm}
\epsfig{file=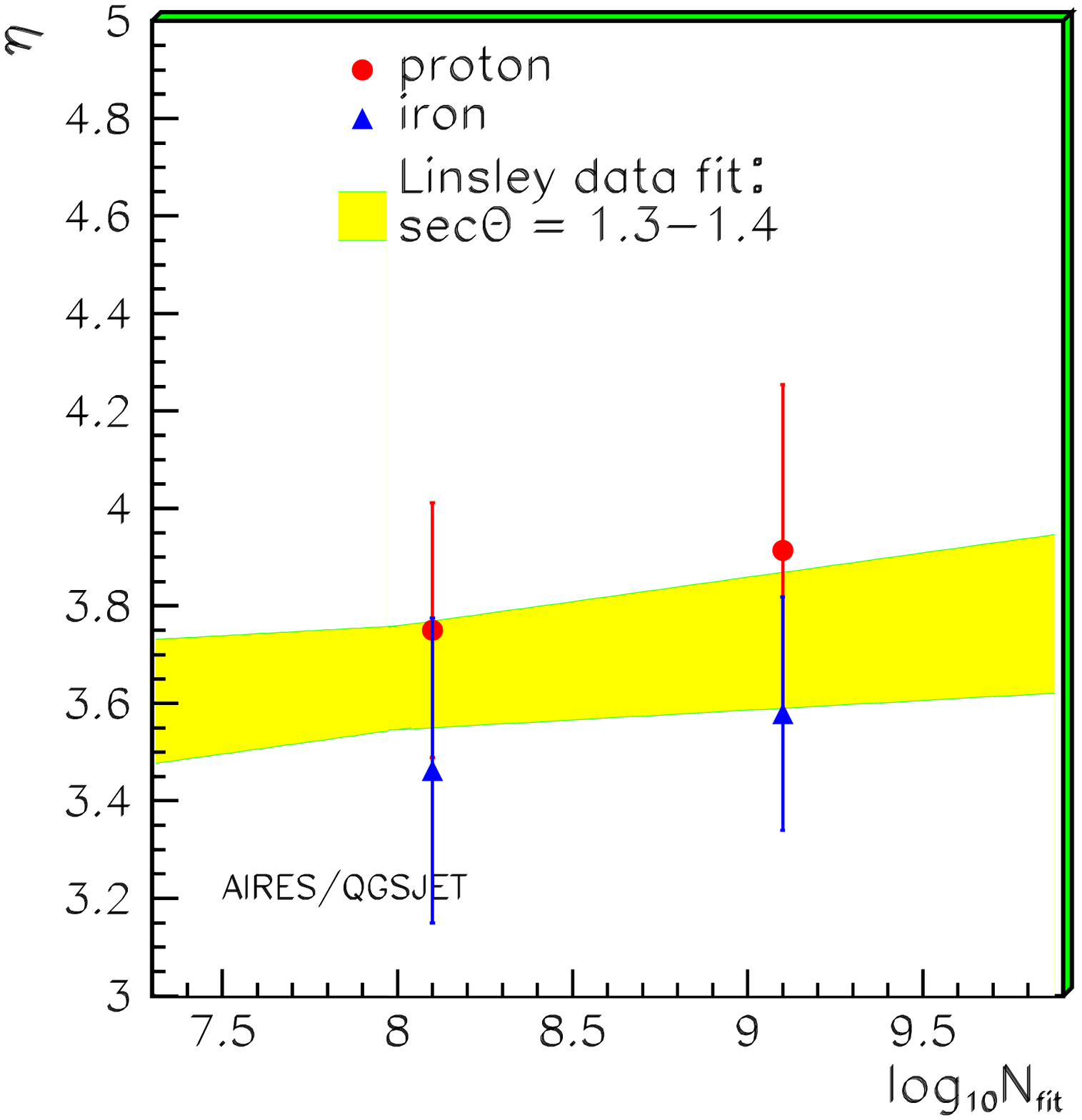,width=8cm}  }
\caption{Comparison of $\eta$ as a function of shower size for $\sec \theta= 1.0 - 1.1$ (left) and $\sec \theta= 1.3 - 1.4$ (right) using {\sc aires/qgsjet98}}
\label{fig:etaN_1}
\end{figure}
\end{center}

\begin{center}
\begin{figure}[p]
\centerline{ 
\epsfig{file=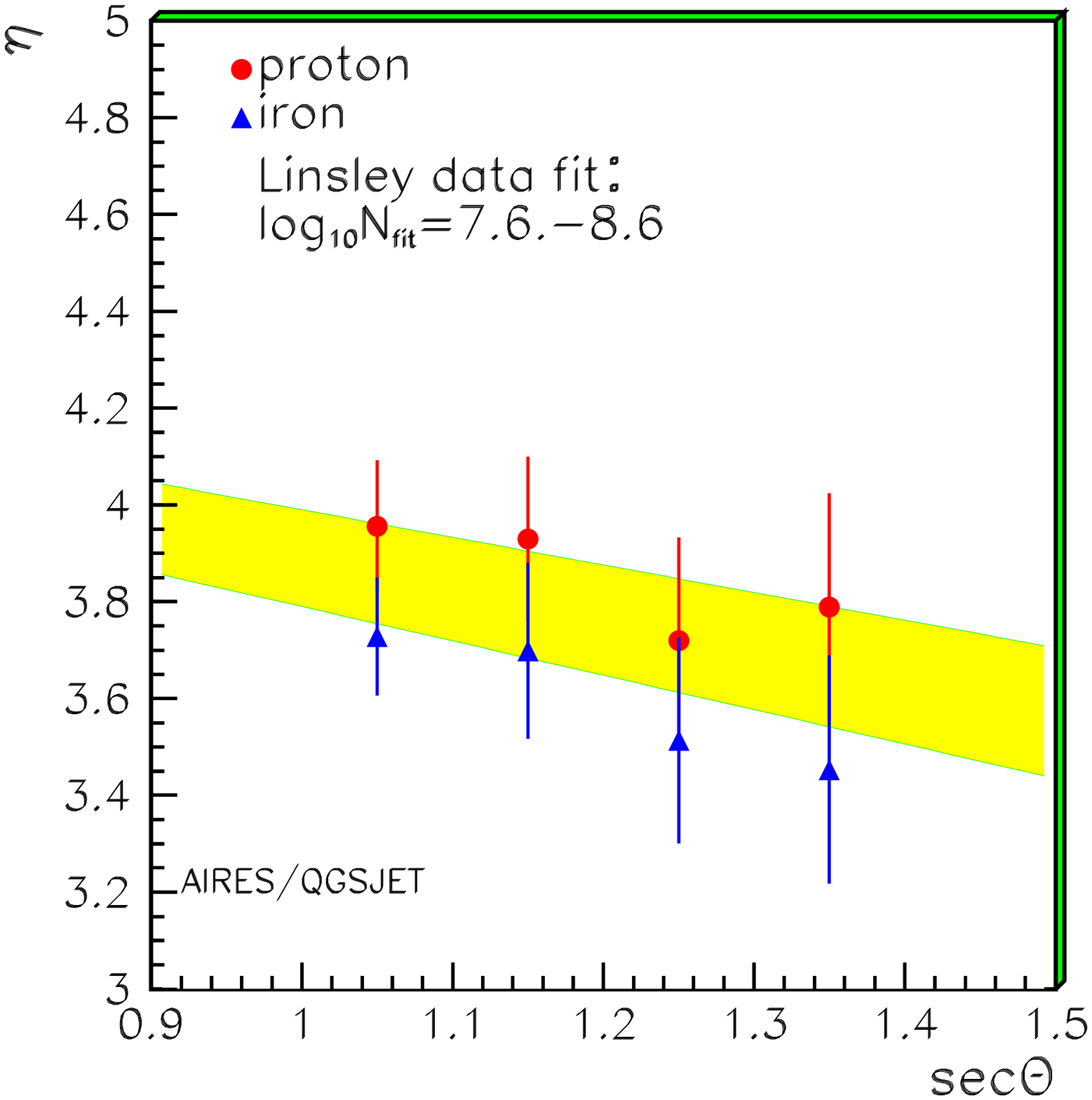,width=8cm}
\hspace{1cm}
\epsfig{file=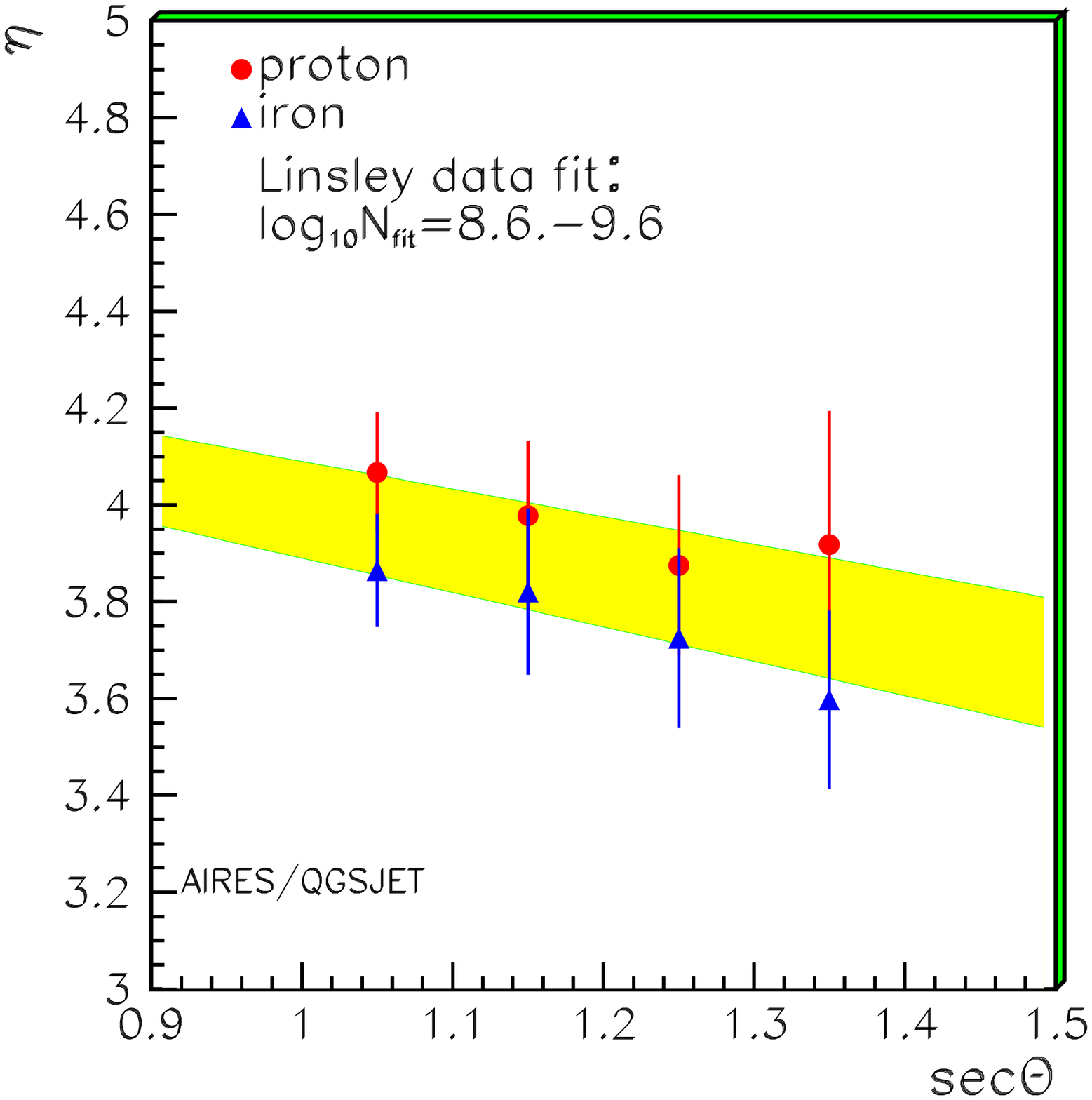,width=8cm}  }
\caption{Comparison of $\eta$ as a function of $\sec \theta$ for the first bin (left) and second bin (right) in $logN_{fit}$}
\label{fig:etasec}
\end{figure}
\end{center}

As a further check, we have calculated  the variation of $\eta $ with shower size 
and zenith angle with Monte Carlo and made comparisons with the Volcano Ranch data.
The number of particles at ground level ($N_{fit}$) is obtained from a fit 
to the lateral distribution function (with $\alpha =1$) for fixed bins of zenith angle.
The variation of $\eta$ with $N_{fit}$ from the calculation has been compared
with the average functional form of $\eta$  given by Equation 3.
The results of this comparison for $\sec \theta =$ 1.0 - 1.1 and 1.3 - 1.4 
can be seen for proton and iron showers in Figure~\ref{fig:etaN_1}. 
The error bars indicate the r.m.s. spread of data which is very much greater 
than the r.m.s. spread of the mean.
The shaded band represents the fit to Volcano Ranch data, 
including the errors given in Equation(3) for $a$ and  $b$.
The variation of $\eta$ with zenith angle is shown in 
Figure ~\ref{fig:etasec} for events with shower size in the range $\log N_{fit}=7.6-8.6$ (left) 
and $\log N_{fit}=8.6-9.6$ (right). 
One can see that the average form of $\eta $ over a realistic range of mass composition,
from proton to iron, is well represented by the simulations.
The error bars represent the r.m.s. spread as before.

\subsection{Fitting VR data mass composition using finite Monte Carlo samples of different primaries.}

We can estimate the primary mass that describes the Volcano Ranch data, assuming a bi-modal composition, 
using a maximum likelihood fit for the best linear combination of pure iron and pure proton samples 
to match the data sample. 
The available data are in bins of $\eta$ [23]: 
the number of data points in several bins is small, so a $\chi^2$ minimization 
is inappropriate. A maximum likelihood technique assuming Poisson statistics was adopted.
The probability of observing a particular number of events $d_i$ in a particular bin is given 
by $\exp^{-f_i}\,f_i^{d_i}/d_i!$  where $f_i$ is the predicted value for the number of events in this particular bin.
If we assume a bi-modal composition of proton and iron with fractions $P_{Fe}$ and $P_p$ then $f_i=C(P_{Fe} + P_p)$ 
where $C$ is the overall normalization factor between numbers of data and Monte Carlo events.
Estimates of the fractions $P_j$ are found by maximizing
 $\ln(L)=\Sigma d_i\ln(f_i)-\ln(d_i!)-f_i$.
Our Monte Carlo samples are at least ten times larger than the data sample
to avoid effects of finite Monte Carlo data size.
\begin{center}
\begin{figure}[p]
\centerline{ 
\epsfig{file=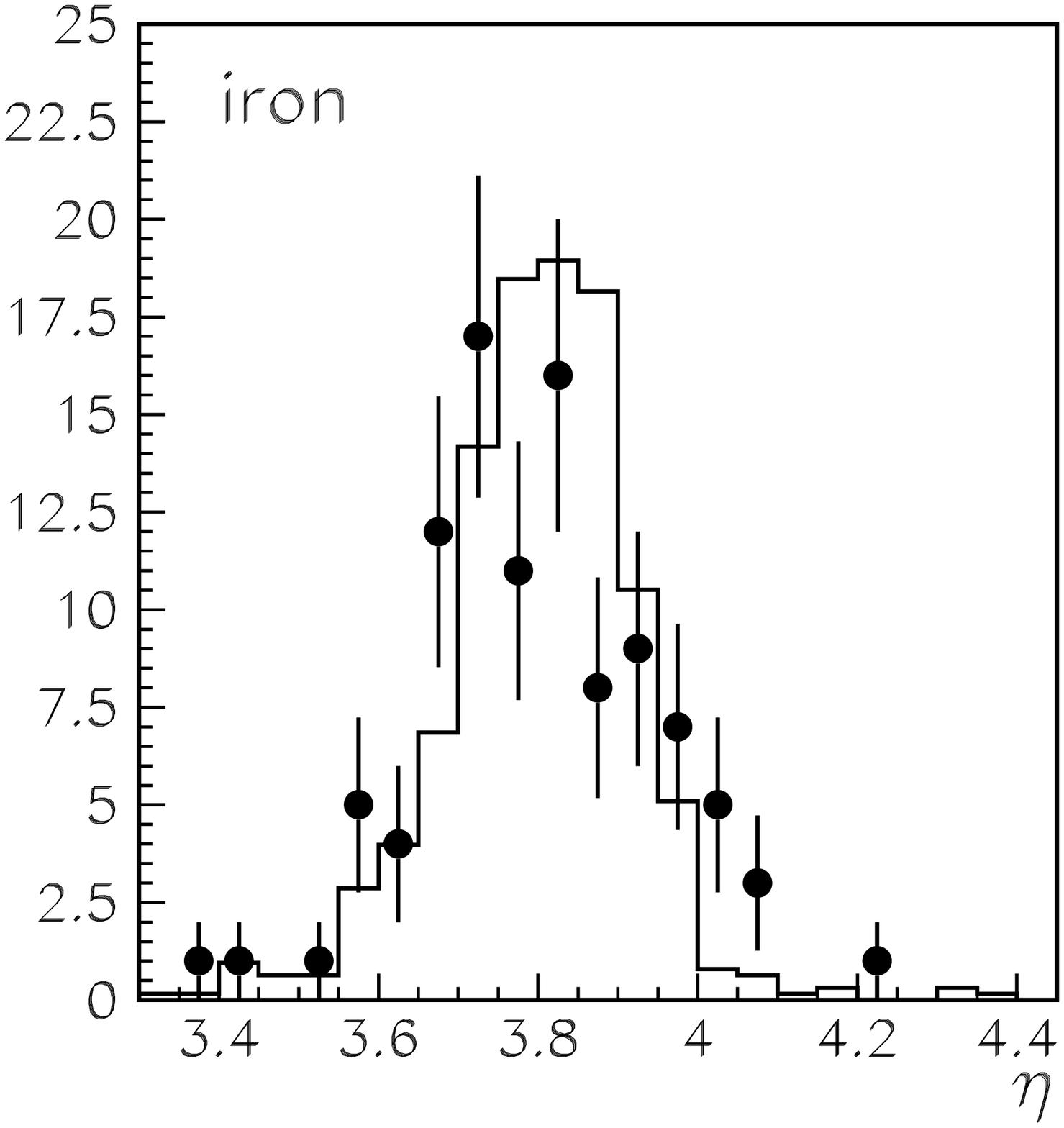,width=8cm}
\hspace{1cm}
\epsfig{file=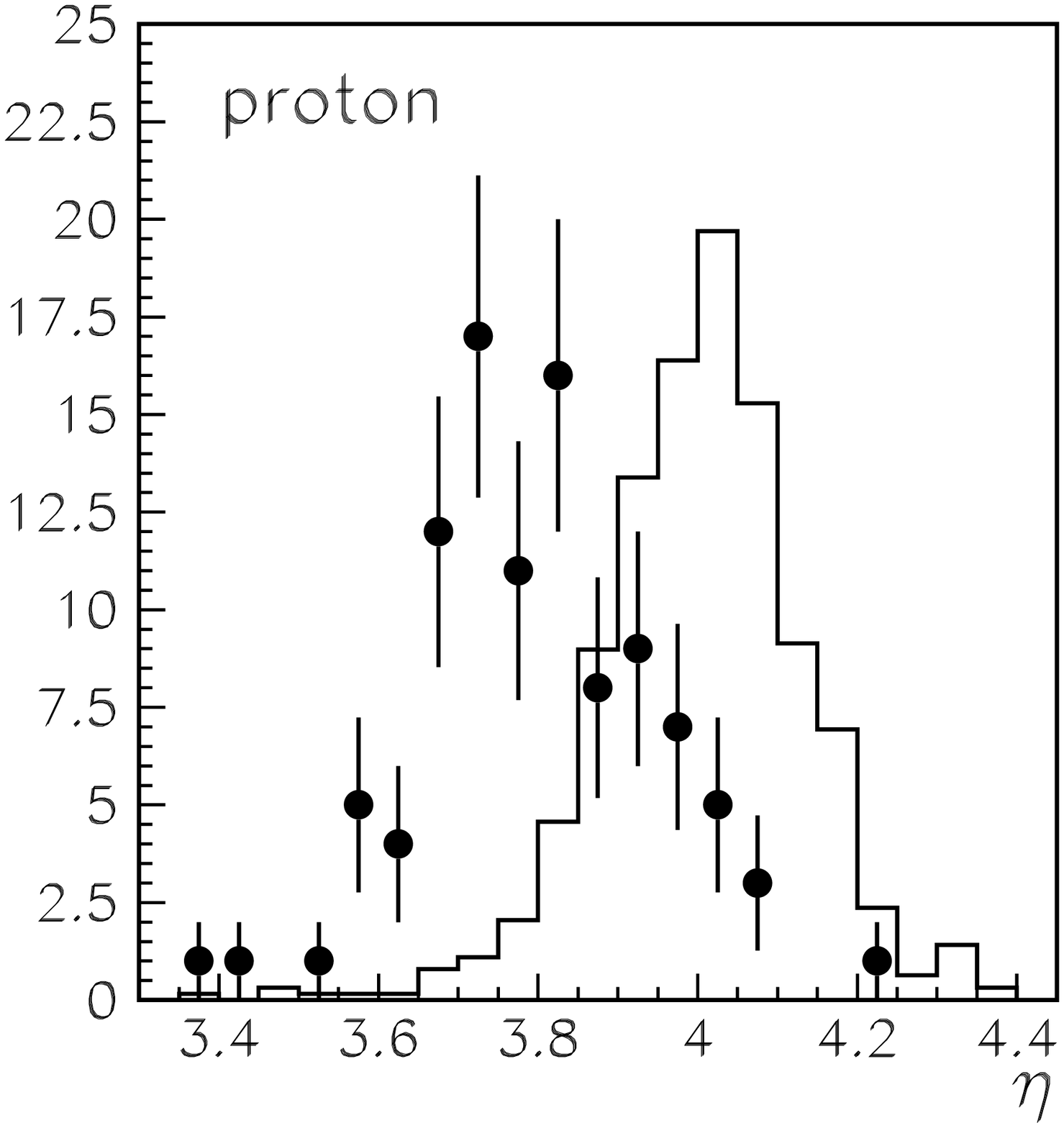,width=8cm}  }
\caption{The measured distributions of $\eta$ (data points) with histograms from Monte Carlo calculations of pure iron (left) and pure proton (right) with $1.0 < \sec\theta < 1.1$, using {\sc qgsjet98}.}
\label{fig:fig1bs}
\end{figure}
\begin{figure}[p]
\epsfig{file=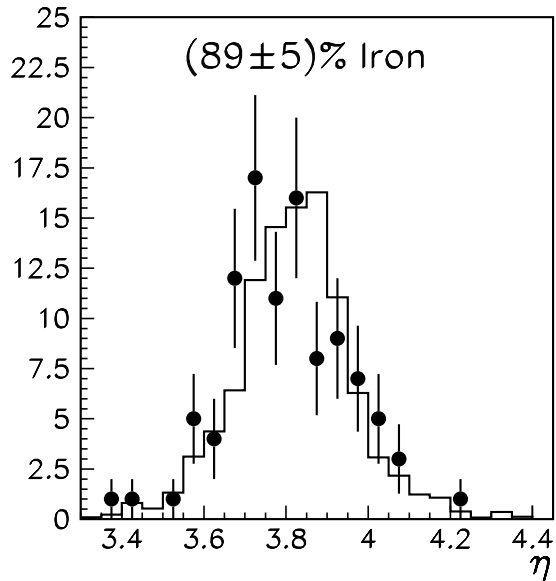,width=8cm} 
\caption{ Composition fit from $\eta$ distribution for the first bin in $1.0 < \sec\theta < 1.1$, using {\sc qgsjet98}. The points are the VR data and the solid line corresponds to the result of the fit.}
\label{fig:fig1b}
\end{figure}
\end{center} 
In Figure~\ref{fig:fig1bs} we compare the Monte Carlo results 
with the Volcano Ranch data points for near vertical showers.  
As can be seen the tail at large $\eta $ in the comparison with iron 
indicates that a lighter component is required to fit the experimental data.  
The best fit gives a mixture with (89 $\pm $ 5) \% of iron, with a corresponding percentage of 
protons, and this distribution of $\eta$ is shown in Figure~\ref{fig:fig1b}.
One detail that Linsley did not describe is the distribution of 
shower sizes that comprise the data set. 
What is recorded is that the median energy was $10^{18}$ eV and that the shower sizes 
are between $4\times10^{7}$ and $6\times10^{9}$.
This corresponds to an energy range approximately between $10^{17}$ eV and $10^{19}$ eV.
To reproduce the data set, a differential energy spectrum of slope -3.5 was chosen for the primary 
energy spectrum.
Simulations show that the whole area of the array was active above  $10^{18}$ eV.
Using this spectrum and the condition that the median energy be $10^{18}$ eV, the calculated 
threshold energy is approximately $10^{17.7}$ eV.
Below $10^{18}$ eV the energy distribution of the data set was smoothed (always under the condition of 
median energy to be $10^{18}$ eV) assuming that the effective area of the detector increases with energy
from the threshold up to $10^{18}$ eV.
The conclusions about mass composition that we presently draw, 
are constrained by uncertainties in 
the details of the energy distribution of events recorded at VR.
Additionally they are constrained by the hadronic model assumed.

The systematic error arising from our lack of knowledge of the energy distribution
of the events has been estimated by repeating the fitting 
procedure with different energy spectra. 
From this analysis we estimate a systematic error of 12\%.
An additional source of systematic error is related to uncertainties 
in the hadronic interaction model: 
following the discussion in~\cite{HaverahPark} 
which relates to the use of {\sc qgsjet98} rather 
than {\sc qgsjet01}, the systematic shift in the fraction of iron is 14\%. 
Showers that are calculated using {\sc qgsjet01} are 
found to develop higher in the atmosphere so that the fraction 
of Fe estimated is reduced from (89 $\pm $ 5)\% to (75 $\pm $ 5)\%.

\begin{center}
\begin{figure}[p] 
\centerline{ 
\epsfig{file=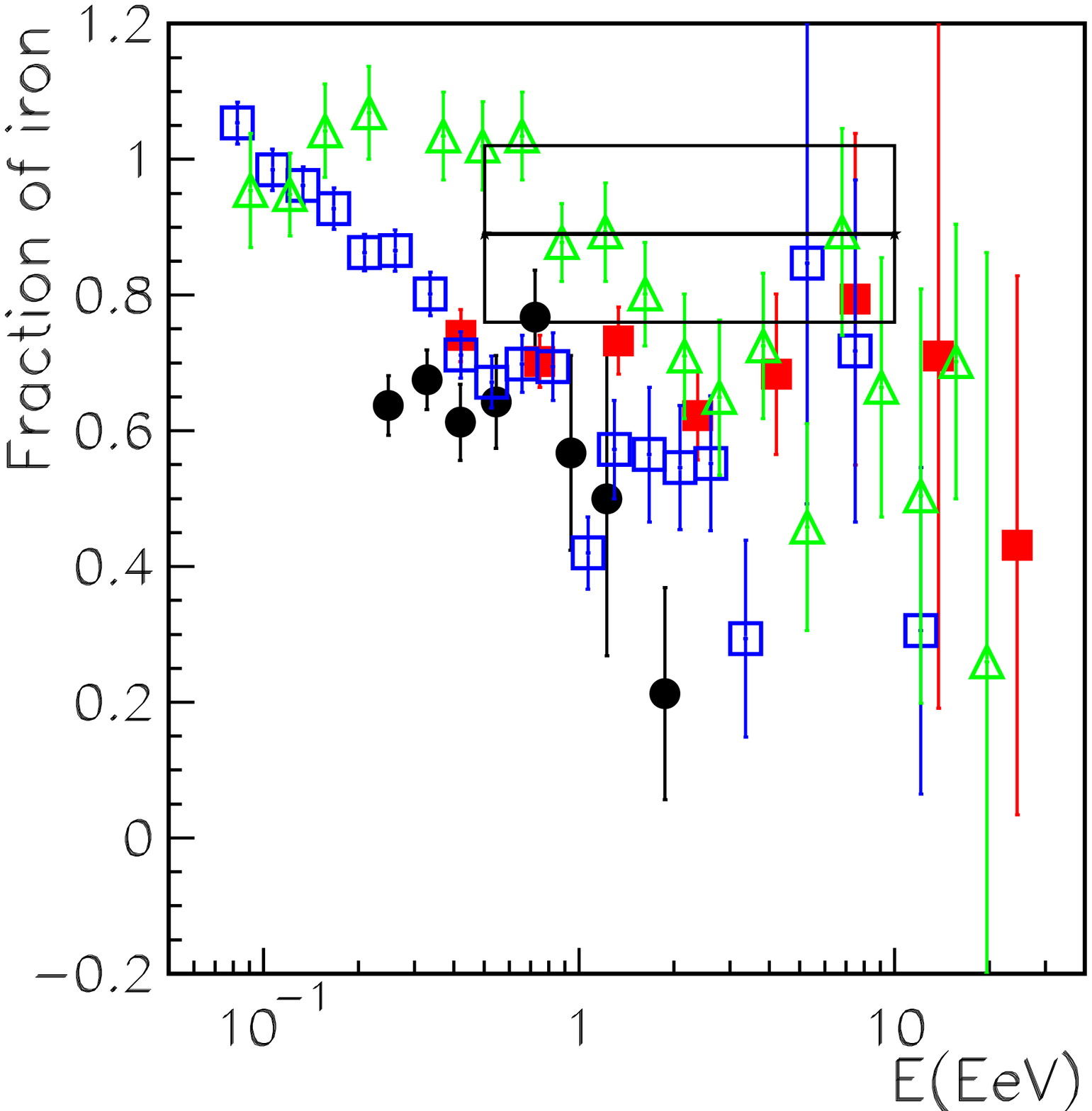,width=8cm}
\hspace{1cm}
\epsfig{file=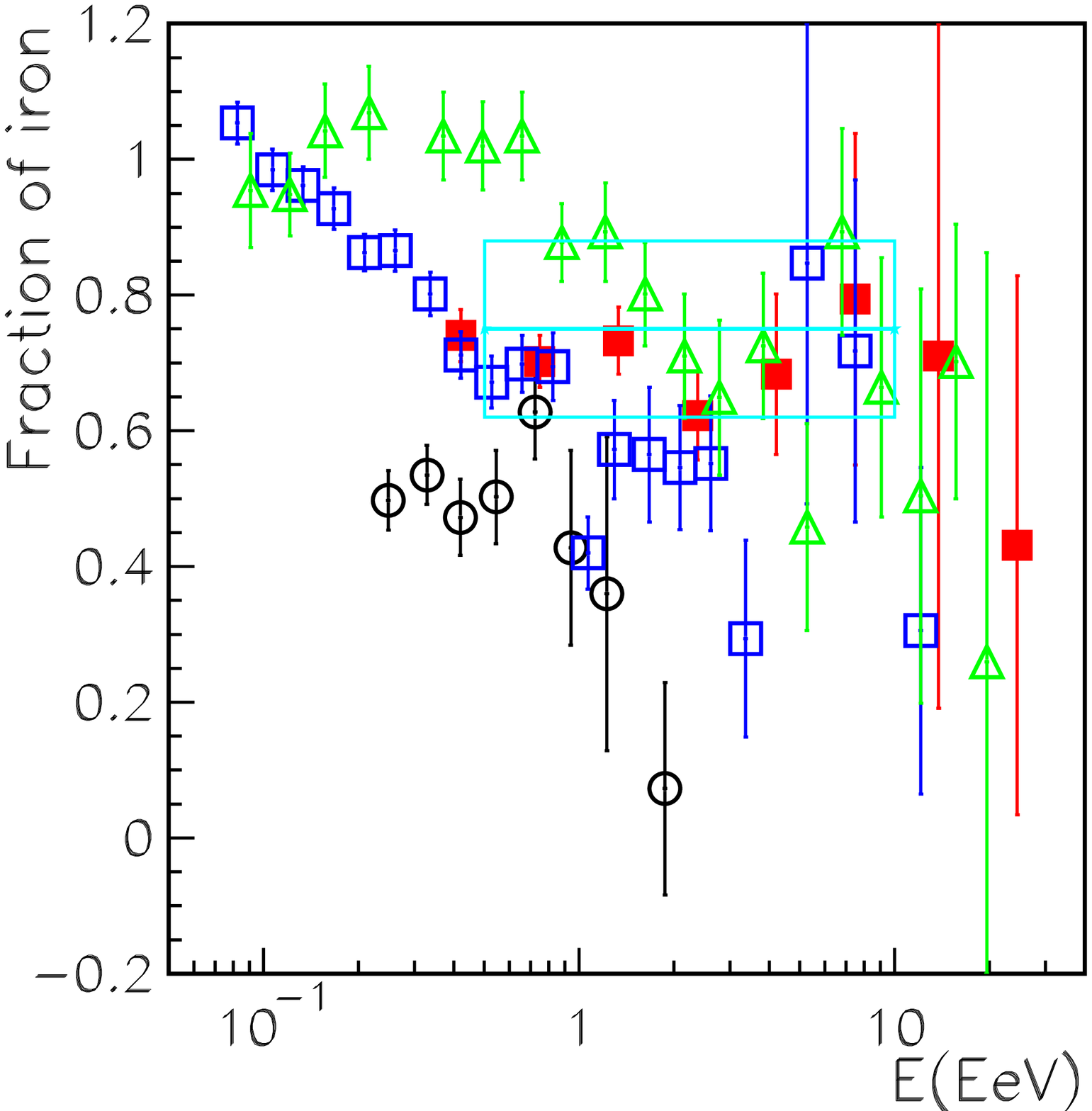,width=8cm}  }
\caption{Fe fraction from various experiments: Fly's Eye ($\triangle$), Agasa A100 ({\tiny $\blacksquare$}), Agasa A1 ({\tiny $\square$}) using {\sc sibyll 1.5} ([4] and references therein) and Haverah Park [1], using {\sc qgsjet98} ({\large $\bullet$})(left) and {\sc qgsjet01} ({\large $\circ$})(right). Mean composition determined in this paper with the corresponding error for the Volcano Ranch energy range using {\sc qgsjet98} (solid line rectangle at left) and an estimation of what it would result using {\sc qgsjet01} following [1] (dashed rectangle at right) is shown.} 
\label{fig:fefrac}
\end{figure}
\end{center}

\section{Comparison with other data} 
In Figure~\ref{fig:fefrac} we present the fraction of iron reported by Fly's Eye, Agasa A100 and A1 (using {\sc sibyll} as the hadronic model) ~\cite{FlyEyeAkeno} and the results for Haverah Park~\cite{HaverahPark} using {\sc qgsjet98} and {\sc qgsjet01}.
Also shown is the mean composition with the corresponding error we get for the Volcano Ranch energy range using {\sc qgsjet98} and an estimation of what we would get using {\sc qgsjet01} as it was done in Haverah Park~\cite{HaverahPark}. 

The inconsistency that exists between several experiments which span different energy ranges,
use different techniques is doubtless enhanced as different hadronic models are used for 
the interpretation of the raw data.
The application of a consistent hadronic model~\cite{FlyEyeAkeno} brings the results of AGASA into better agreement with the Fly's Eye conclusion, while their original analysis stated that there is no indication of changing composition. 
In Figure~\ref{fig:fefrac} the iron fraction for Fly's Eye and AGASA 
corresponds to the re-interpretation under the {\sc sibyll} 
hadronic model, including triggering efficiency effects; 
the original analysis from AGASA is not presented here.

While Haverah Park, Volcano Ranch and Akeno-AGASA infer $X_{max}$, and hence the overall composition, from properties 
of secondary particle distribution at ground, Fly's Eye and HiRes experiments observe a image of the longitudinal shower profile and 
derive $X_{max}$ directly. 
Nonetheless the estimates can be biased due to the poor knowledge of 
atmospheric properties as recent studies of atmospheric profiles have suggested
~\cite{nuevo}.

\section{Conclusions}
Measurements of the steepness of the lateral distribution $\eta $ were made at Volcano Ranch 
on a shower-to-shower basis for fixed bins of zenith angle.
We have compared the measured distribution of $\eta $ to our Monte Carlo results for proton and iron primaries 
using {\sc qgsjet98} including the scintillator response of the detectors in the Volcano Ranch array.
Our ability to reproduce Volcano Ranch lateral distribution measurements give us confidence that our 
analysis procedure is correct.

The cosmic ray mass composition, deduced from Volcano Ranch data, is compatible
with mean fraction (89 $\pm$ 5(stat) $\pm$ 12(sys)) \% of iron in a bi-modal proton and iron mix, 
in the whole energy range $10^{17.7}$ eV to $10^{19}$ eV, mean energy  $10^{18}$ eV. 
Following the discussion in ~\cite{HaverahPark}, we estimate that this fraction would be
reduced to 75 \%, with the same {\sc qgsjet01} model adopted.

As mentioned above, some experiments find indications that the composition is becoming lighter with energy. 
However, in a recent work using Haverah Park data above $10^{19}$ eV,
it was claimed that the observed time structure of the shower front 
can be best understood if iron primaries are dominant here ~\cite{maria}. 
The differences between measurements on mass composition needs to be addressed further 
if more solid conclusions on the origin, acceleration or propagation of cosmic rays 
are to be reached. 
The rate of change with energy and the average mass are still under debate.

{\acknowledgements}
We warmly thank the late John Linsley for the discussions that we had 
about his data during the 27th ICRC in Hamburg, and for his permission to proceed
with an analysis along the lines described above. However, the
interpretation given is that of the authors alone, as it was not possible to
discuss the outcome of the work with Linsley before his death. 
Work at Universidad de La Plata is supported by ANPCyT (Argentina), 
at Northeastern University by the U.S. National Science Foundation
and at the University of Leeds by PPARC, UK (PPA/Y/S/1999/00276). TPM gratefully acknowledges the support of the American Astronomical Society in the form of a travel grant to attend the 28th ICRC, Japan. MTD thanks the John Simon Guggenheim  Foundation for a fellowship.\\

\end{document}